\documentclass[12pt]{article}

\usepackage{amsfonts,amsmath,amssymb,graphicx,subfigure,a4wide,wrapfig,pifont}

\graphicspath{{./figures/}}

\begin{document}

\title{\bf{Neutrino in magnetic fields: from the first studies
to the new effects in neutrino oscillations}}

\author{
 Alexander Studenikin
 \\
 e-mail:
  studenik@srd.sinp.msu.ru
  \\
Department of Theoretical Physics,
  \\
Moscow State University,
  \\
  119992  Moscow, Russia}
  \date{}
  \maketitle
\begin{abstract}

In this paper I should like to present  {\bf { {the four new
effects}}} in neutrino oscillations that have been recently
investigated in my research group at the Department of Theoretical
Physics of the Moscow State University. Due to the fact that these
studies were stimulated by our previous  research of neutrino
interactions in the presence of magnetic fields, and also because
the year 2004 commemorates the 40th years jubilee since the the
first paper on the neutrino interaction in a magnetic field was
published, a short review on the first papers dedicated to the
problem of neutrinos in magnetic fields, and also on the recent
results in this field, prefaces (Section 1) the discussion on {\bf
the new effects} in neutrino oscillations. Section 2 is devoted to
our recent studies of the electromagnetic properties of  a {\it
massive} neutrino, including the neutrino magnetic moment for
different values of neutrino mass. In Section 3 we discuss {\it
\bf the four new effects} in neutrino spin and flavour
oscillations in different background environments.
\end{abstract}

\section{  Beta-decay of neutron in magnetic field}

About forty years ago the first studies of the neutrino
interaction in the presence of a magnetic field were performed in
the two papers \cite{Kor64TerLysKor65} dedicated to the polarized
neutron beta-decay $n\rightarrow p+e+\tilde\nu_{e}$ in a magnetic
field. In these papers the probability of the polarized neutron
beta-decay in the presence of a magnetic field was derived, as
well as the asymmetry in the neutrino emission was studied for the
first time. It was shown that the differential rate of the process
exhibits the resonance spikes which appears, for the given
magnetic field strength, each time when the final electron energy
is exactly equal to one of the allowed Landau energies in the
magnetic field. It was also shown that the total rate depends on
the initial neutron polarization, in contrast with the field-free
case when the neutron polarization dependence disappears from the
rate during integration over the phase space of the process. The
range of magnetic field strengths considered in these papers span
up to subcritical fields $B\geq B_{0}={m^{2} \over e}= 4.41 \times
10^{13} \ Gauss$. It worth to be noted here that these studies
were performed before the discovery of pulsars
\cite{HewBelPilScoCol68} where such a strong magnetic fields are
believed to exist.

In the two well known papers \cite{MatOCon69Fas69}, published a
few years later, the results of \cite{Kor64TerLysKor65} for the
neutron decay rate in a magnetic field were re-derived and the
problem of the neutron star cooling was considered. However there
were no discussion on the asymmetry in neutrino emission in the
papers of ref.\cite{MatOCon69Fas69}.

Very strong magnetic fields are also supposed to exist in the
early Universe (for a recent review see, e.g. \cite{GraRub01}). As
it was discussed for the first time in \cite{Gre69MatOCon70}, the
weak reaction rates of the processes like
\begin{equation} \label{URCA}
  n\rightarrow p + e +\tilde\nu_e, \ \
  \nu_e + n\leftrightharpoons e+p, \ \
  p+\tilde\nu_e \leftrightharpoons n+e^{+},
\end{equation}
which determine the iter-conversion between neutrons and protons
and set the $n/p$-ration in various environments, can be
significantly modified under the influence of magnetic fields and,
as a consequence, influence the primordial nucleosynthesis
affecting production of $^{4}{He}$ .

The aforementioned studies of neutrino interactions in the
presence of magnetic fields
\cite{Kor64TerLysKor65,MatOCon69Fas69,Gre69MatOCon70} gave the
birth to {\bf the neutrino astrophysics in magnetic fields}.

Various authors (the first papers in this field are given in ref.
\cite{Chu84DorRodTer84-85ZahLos85}) argued that asymmetric
neutrino emission in the direct URCA processes (\ref{URCA}) during
the first seconds after the massive star collapse could provide
explanations for the observed pulsar velocities. We have shown
\cite{Stu88} that in order to get a correct prediction for the
direction and value of the kick velocity of a pulsar one has to
account not only for the amount of radiated in the processes
(\ref{URCA}) neutrinos but also for the fact that the average
momentum of neutrinos propagating in the opposite directions are
not equal one to each other. Some of the recent studies in this
field can be found in
\cite{DuaQia0401634LeiPer98Goy98LaiQia98ArrLai99GvoOgn99}. A lot
of other different mechanisms for the asymmetry in the neutrino
emission from a magnetized pulsar has been also studied previously
(see e.g. \cite{KusSeg96BisKog97AkhLanSci97LaiQia98}). For more
complete references to the performed studies on the neutrino
mechanisms of the pulsar kicks see the second review paper of
ref.\cite{Rou97BhaPal02}.

Recently we have developed \cite{ShiStu0402154} the relativistic
approach to the inverse $\beta$-decay of a polarized neutron, $\nu
_{e} + n \rightarrow p + e ^{-}$, in a magnetic field. This
process can be also important for the neutrino transport inside
the magnetized pulsar and contribute to the kick velocities
\cite{Rou97BhaPal02}. As we have shown \cite{ShiStu0402154}, in
strong magnetic fields the cross section can be highly anisotropic
in respect to the neutrino angle. In the particular case of
polarized neutrons, matter becomes even transparent for neutrinos
if neutrinos propagate against the direction of neutrons
polarization.


\section{ Electromagnetic properties of a massive neutrino
\cite{DvoStuPRD04,DvoStuJETP04}}
\date{}

It is well known that neutrino with non-zero mass has non-trivial
electromagnetic properties. In particular, the Dirac massive
neutrino can posses non-vanishing magnetic and electric dipole
moments. Noted here that the massive Majorana neutrino can't have
neither magnetic no electric moments in vacuum. However, the
Majorana neutrino can have flavour non-diagonal ( transition )
magnetic and electric moments. It is believed that non-zero
neutrino magnetic moment could have an important impact on
astrophysics and cosmology.

It is also well known \cite{FujShr80} that in the minimally
extended Standard Model with $SU(2)$-singlet right-handed neutrino
the one-loop radiative correction generates neutrino magnetic
moment which is proportional to the neutrino mass
\begin{equation}
\mu_{\nu}={3 \over {8 \sqrt{2}\pi^{2}}}eG_{F}m_{\nu}=3 \times
10^{- 19}\mu_{0}\bigg({m_{\nu} \over {1 \mathrm{eV}}}\bigg),
\end{equation}
where $\mu_{0}=e/2m$ is the Bohr magneton, $m_{\nu}$ and $m$ are
the neutrino and electron masses. There are also models
\cite{KBMVVFY76-87} in which much large values for magnetic
moments of neutrinos are predicted. So far, the most stringent
laboratory constraints on the electron, muon, and tau neutrino
magnetic moments come from elastic neutrino-electron scattering
experiments: $\mu_{\nu_{e}} \leqslant 1.5 \times 10^{-10} \mu_{0},
\ \ \mu_{\nu_{\mu}} \leqslant 6.8 \times 10^{-10} \mu_{0},\ \
\mu_{\nu_{\tau}} \leqslant 3.9 \times 10^{-10} \mu_{0},$
\cite{RevParPhy02}.

  Recently we have considered \cite{DvoStuPRD04,DvoStuJETP04}
the massive Dirac neutrino electromagnetic form factors in the
context of the standard model supplied with
$\mathrm{SU}(2)$-singlet right-handed neutrino. Using the
dimensional-regularization scheme, we have performed the most
general study of the {\it massive} neutrino one-loop vertex
function in the general $R_{\xi}$ gauge  exactly accounting for
the masses of all particles in polarization loops. In
\cite{DvoStuJETP04} we have also accounted for neutrino mixing
effects and discussed  the massive neutrino anapole moment in
details.

\begin{wrapfigure}{l}{7cm}
  \centering
  \includegraphics[scale=.35]{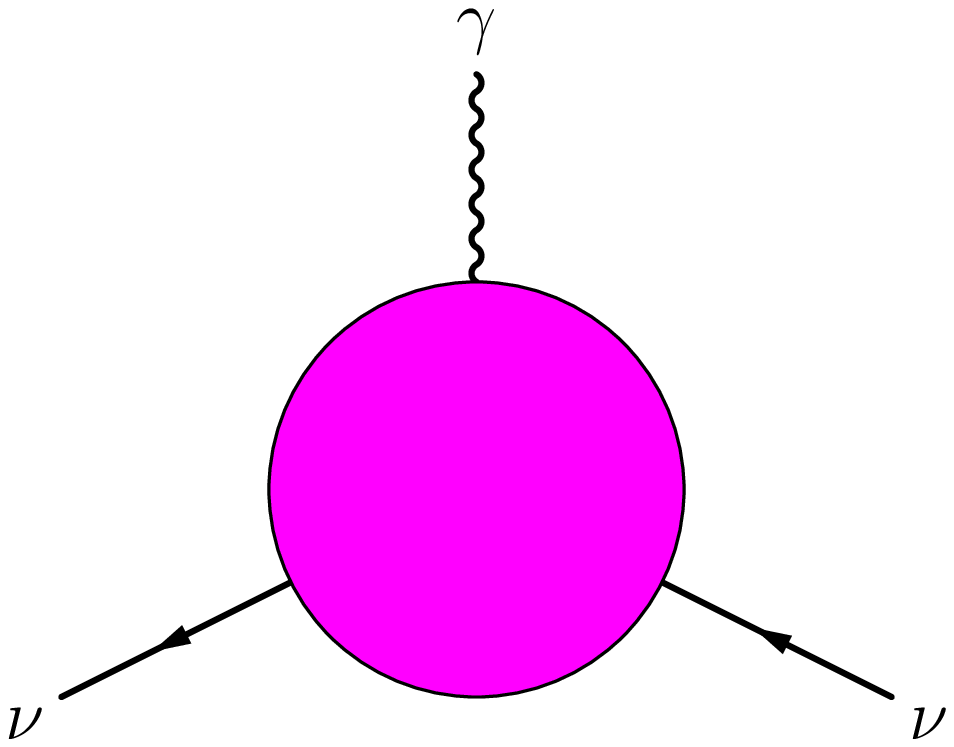}
  \caption{Neutrino electromagnetic vertex function}
  \label{nuvert}
\end{wrapfigure}

The neutrino electromagnetic vertex function $\Lambda_{\mu}(q)$ is
shown in Fig.~\ref{nuvert}.

\begin{figure}
  \centering
  \subfigure[]
  {\label{prverta}
    \includegraphics[scale=.7]{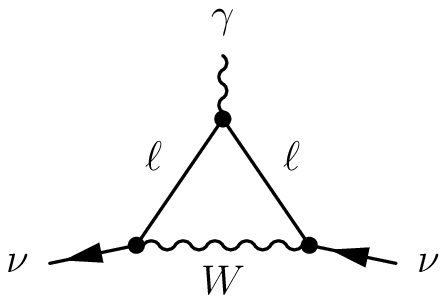}}
    \hspace{.5cm}
  \subfigure[]
  {\label{prvertb}
  \includegraphics[scale=.7]{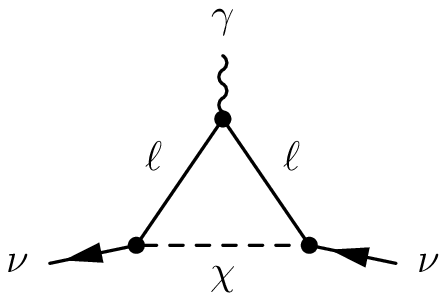}}
    \hspace{.5cm}
  \subfigure[]
  {\label{prvertc}
  \includegraphics[scale=.7]{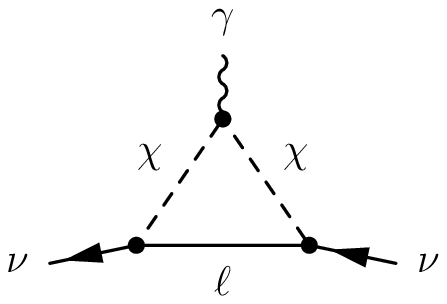}}
    \\
  \subfigure[]
  {\label{prvertd}
  \includegraphics[scale=.7]{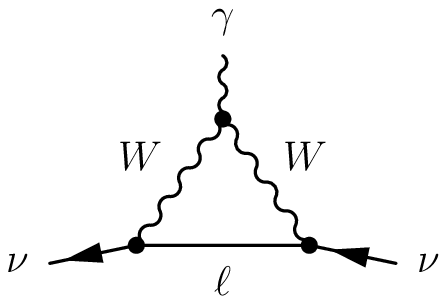}}
    \hspace{.5cm}
  \subfigure[]
  {\label{prverte}
  \includegraphics[scale=.7]{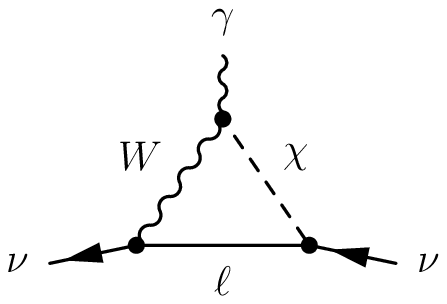}}
    \hspace{.5cm}
  \subfigure[]
  {\label{prvertf}
  \includegraphics[scale=.7]{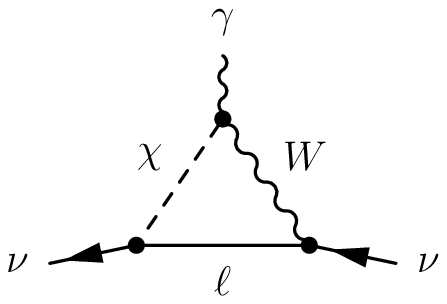}}
    \caption{\subref{prverta}-\subref{prvertf} proper vertices.}
\end{figure}

\begin{figure}
  \centering
  \subfigure[]
  {\label{gZverta}
    \includegraphics[scale=.7]{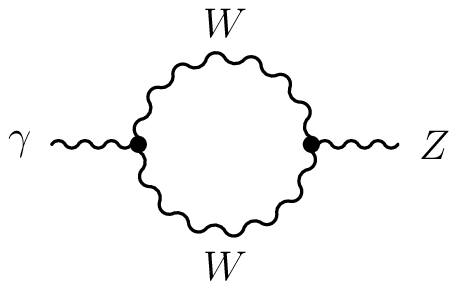}}
    \hspace{.5cm}
  \subfigure[]
  {\label{gZvertb}
  \includegraphics[scale=.7]{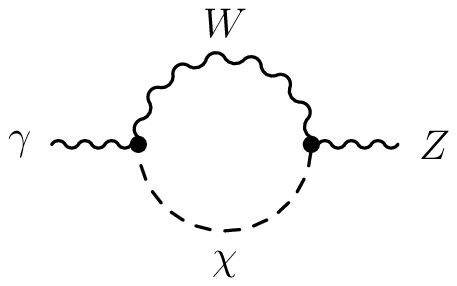}}
    \hspace{.5cm}
  \subfigure[]
  {\label{gZvertc}
  \includegraphics[scale=.7]{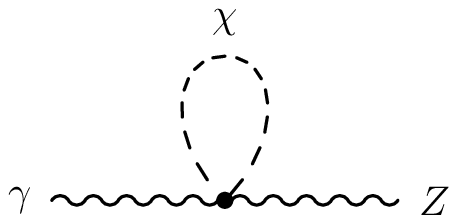}}
    \hspace{.5cm}
  \subfigure[]
  {\label{gZvertd}
  \includegraphics[scale=.7]{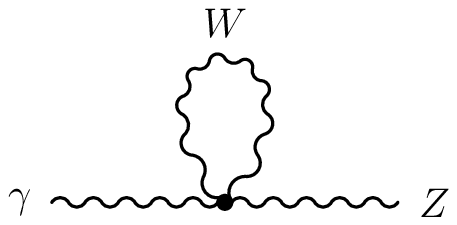}}
    \\
  \subfigure[]
  {\label{gZverte}
  \includegraphics[scale=.7]{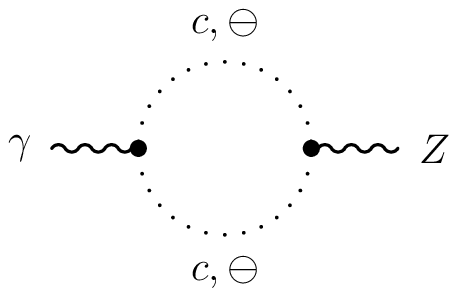}}
    \hspace{.5cm}
  \subfigure[]
  {\label{gZvertf}
  \includegraphics[scale=.7]{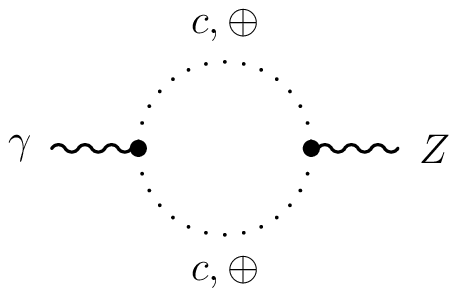}}
    \hspace{.5cm}
  \subfigure[]
  {\label{gZvertg}
  \includegraphics[scale=.7]{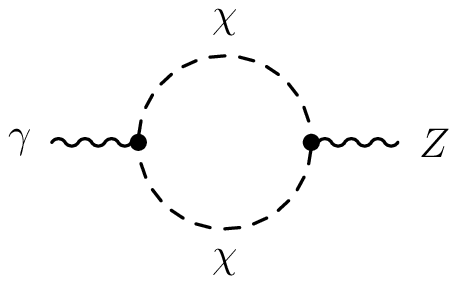}}
    \hspace{.5cm}
  \subfigure[]
  {\label{gZverth}
  \includegraphics[scale=.6]{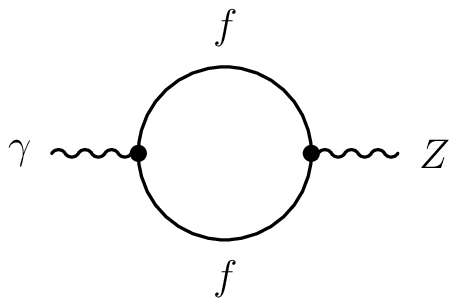}}
    \caption{\subref{gZverta}-\subref{gZverth}
    The $\gamma -Z$ self-energy diagram.
    $f$ denotes the electron, muon, and $\tau$-lepton as well as $u$, $c$ ,$t$,
    $d$, $s$, and $b$ quarks.}
\end{figure}

 The one-loop contributions to the neutrino electromagnetic
vertex $\Lambda_{\mu}(q)$ are given by the two types of Feynman
diagrams: the proper vertices [Fig.~\ref{prverta}-\ref{prvertf}]
and the $\gamma-Z$ self-energy diagrams
[Fig.~\ref{gZverta}-\ref{gZverth}].

The matrix element of the electromagnetic current between neutrino
states can be presented in the form

\begin{equation*}
  \langle{\nu}(p^{\prime})|J_{\mu}^\mathrm{EM}|\nu(p)\rangle=
  \bar{u}(p^{\prime})\Lambda_{\mu}(q)u(p),
\end{equation*}
where the most general expression for the electromagnetic vertex
function $\Lambda_{\mu}(q)$ reads
\begin{equation*}
  \Lambda_{\mu}(q)=
  f_{Q}(q^{2})\gamma_{\mu}+f_{M}(q^{2})i\sigma_{\mu\nu}q^{\nu}-
  f_{E}(q^{2})\sigma_{\mu\nu}q^{\nu}\gamma_{5}+
  f_{A}(q^{2})(q^{2}\gamma_{\mu}-q_{\mu}{\not q})\gamma_{5}.
\end{equation*}
Here $f_{Q}(q^{2})$, $f_{M}(q^{2})$, $f_{E}(q^{2})$ and
$f_{A}(q^{2})$ are respectively the electric, dipole electric,
dipole magnetic, and anapole neutrino form factors,
$q_{\mu}=p^{\prime}_{\mu}-p_{\mu}$,
$\sigma_{\mu\nu}=(i/2)[\gamma_{\mu},\gamma_{\nu}]$,
$\gamma_5=-i\gamma^0\gamma^1\gamma^2\gamma^3$.

\begin{wrapfigure}{r}{5.0cm}
  \centering
  \includegraphics[scale=.7]{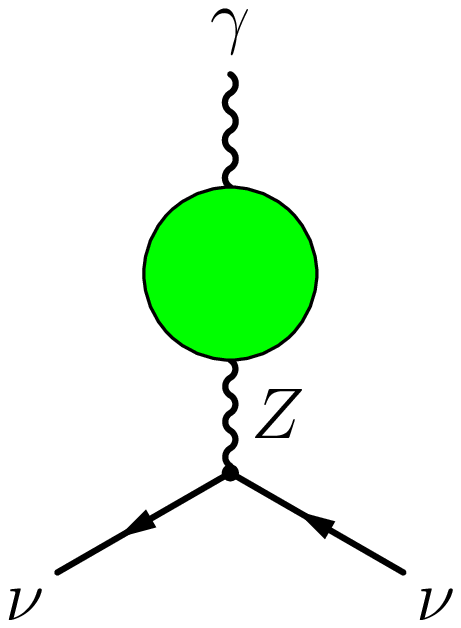}
  \caption{The contributions of $\gamma-Z$ diagrams}
  \label{gZcontr}
\end{wrapfigure}

The contributions of the $\gamma-Z$ diagrams are schematically
presented in Fig.~\ref{gZcontr}.
Performing the direct calculations of the loop diagrams we
established that:

1) neutrino electromagnetic function consist of only three
electromagnetic form factors ( in the case of a model with $CP$
conservation), 2) there is such a gauge in which all
electromagnetic form factors are finite, i.e. they do not contain
ultraviolet divergences. The values of the gauge fixing parameters
are
\begin{equation}
  \alpha_W = \frac{1}{9}(138+151\tan^2\theta_W), \ \
  \alpha_Z = +\infty
\end{equation}

We also found \cite{DvoStuPRD04,DvoStuJETP04} the closed integral
expressions for electric, magnetic, and anapole form factors of a
{\it massive} neutrino. On this basis we have derived the electric
charge (the value of the electric form factor at zero momentum
transfer), magnetic moment, and anapole moment of a {\it massive}
neutrino. By means of the direct calculations for the case of a
{\it massive} neutrino we have shown that: 1) electric charge is
independent of the gauge parameters and is equal to zero, 2)
magnetic moment does not depend on the choice of the
  gauge.

The recent LEP data require that the number of light neutrinos
coupling to Z boson is exactly three, whereas any additional
neutrino, if this particle exist, must be heavy. That is the
reason to consider the neutrino magnetic moment for various ranges
of particles masses. We have obtained \cite{DvoStuPRD04} the
values of the neutrino magnetic moment for light (for this
particular case see also \cite{FujShr80,CarBerVidZep00}),
intermediate and heavy massive neutrino:
\begin{equation*}
  1)\ \  m_\nu\ll m_\ell\ll M_W,
  \end{equation*}
  \begin{equation*}
    \mu_{\nu}=
    {\frac{eG_{F}}{4\pi^{2}\sqrt{2}}}m_{\nu}
    {\frac{3}{4(1-a)^{3}}}
    (2-7a+6a^{2}-2a^{2}\ln a-a^{3}),  \ \ a=\big( {m_{l} \over
    M_{W}}\big)^{2},
  \end{equation*}
  \begin{equation*}
  2) \ \ m_\ell\ll m_\nu\ll M_W,
\end{equation*}
  \begin{equation*}
    \mu_{\nu}={\frac{3eG_{F}}{8\pi^{2}\sqrt{2}}}m_{\nu}
    \left\{
    1+{\frac{5}{18}}b
    \right\}, \ \ b=\big( {m_{\nu} \over
    M_{W}}\big)^{2},
  \end{equation*}
\begin{equation*}
  3) \ \ m_\ell\ll M_W\ll m_\nu,
\end{equation*}
  \begin{equation*}
    \mu={\frac{eG_{F}}{8\pi^{2}\sqrt{2}}}m_{\nu}.
  \end{equation*}

In the conclusion of this section, we should like to note that the
neutrino electromagnetic properties are affected by the external
environment. In particular, a neutrino can acquire an electric
charge in a magnetized matter \cite{OraSemSmo86}, also the value
of the neutrino magnetic moment can be significantly shifted by
the presence of strong external magnetic fields
\cite{BorZhuKurTer85,EgoLikStuEPP99}. The recent study of the
neutrino electromagnetic vertex in a magnetized matter can be
found in \cite{Nie03}.





\section{ Four new effects in neutrino oscillations in
background environments
\cite{LikStu95,EgoLobStuPLB00,DvoStuYF01_04,LobStuPLB01,
DvoStuJHEP02,GriLobStuPLB02,LobStuPLB03,StuYF04, DvoGriStu04}}

\date{}

In this section we present {\it \bf the four new effects} in
neutrino oscillations that have been recently studied in our
papers \cite{LikStu95,EgoLobStuPLB00}, \cite{DvoStuYF01_04}-
\cite{DvoStuJHEP02}, and \cite{GriLobStuPLB02}-
\cite{DvoGriStu04}.

The thorough research on neutrino oscillations were launched by
the pioneering paper of Bruno Pontecorvo \cite{Pon57}. Indeed, it
is surprising that after more than 45 years of activity in this
field (for some of the most important papers see
refs.\cite{MakNakSakPTP62,GriPonPLB69_BilPonPLB76,Wol78,MikSmi85,
SchVall81VolVysOku86LimMar88Akh88VidWud90Smi91AkhPetSmi93}) it is
still possible to discover new effects.

  The whole story began several years ago  when we
studied neutrino spin oscillations in strong magnetic fields
\cite{LikStu_JETP95} and to our much surprise realized that at
that time in literature there were no even attempts to consider
neutrino spin oscillations in any electromagnetic field
configuration rather than constant in time and transversal in
respect to neutrino motion magnetic fields (see, for example,
\cite{SchVall81VolVysOku86LimMar88Akh88VidWud90Smi91AkhPetSmi93}).
Further more, in all of the studies of neutrino spin and also
flavour oscillations in matter, performed before 1995, the matter
effect was treated only in the non-relativistic limit (matter was
always supposed to be slowly moving or to be at rest, see for
example \cite{Wol78}).

The first our attempt \cite {LikStu95} to consider neutrino
flavour oscillations in matter in the case when matter is moving
with relativistic speed was made in 1995. In that our study we
have tried to apply the Lorentz invariant formalism to describe
neutrino flavour oscillations and realized that the value of the
matter term in the neutrino effective potential can be
significantly changed if matter is moving with relativistic speed.
In particular, in  \cite{LikStu95} we have shown that the
difference of neutrino effective potentials, $V_{eff}$,   in
non-polarized relativistically moving matter composed of electrons
is proportional to $V_{eff}\sim (1-\vec\beta \vec v_{e})$, where
$\vec\beta$ and $\vec v_{e}$ are the speeds of neutrino and
electrons, correspondingly. Thus, we have observed in 1995 that,
for the case of matter moving with relativistic speed along the
direction of the ultra relativistic neutrino propagation, the
effect of matter in oscillations is washed out.

We have continued \cite{EgoLobStuPLB00} our studies on evaluation
of the Lorentz invariant formalism in neutrino oscillations in
1999 when have developed an approach that enables us to consider
neutrino spin oscillations in arbitrary electromagnetic field
configurations. We have shown \cite{EgoLobStuPLB00} that the
Bargmann-Michel-Telegdi equation \cite{BarMicTel59}, describing a
neutral particle spin evolution under the influence of an
electromagnetic field, can be generalized for the case of a
neutrino moving in electromagnetic fields and matter by
implementing the substitution  of the external electromagnetic
field tensor, $F_{\mu \nu}=(\vec E, \vec B)$, according to the
prescription $F_{\mu\nu} \rightarrow F_{\mu\nu}+G_{\mu\nu}$. The
anti-symmetric tensor $G_{\mu \nu}=(- \vec P, \vec M)$ can be
constructed with use of the neutrino speed, matter speed, and
matter polarization four-vectors under the natural assumptions
that the neutrino spin evolution equation have to be linear over
$F_{\mu \nu}$ and over the other mentioned above vectors. From
this new generalized BMT equation for the neutrino spin evolution
in an electromagnetic field and matter we have finally arrive to
the following equation for the evolution of the
three-di\-men\-sio\-nal neutrino spin vector $\vec S $:

\begin{equation}\label{S} {d\vec S
\over dt}={2\mu \over \gamma} \Big[ {\vec S \times ({\vec
B_0}+\vec M_0)} \Big],
\end{equation}
\begin{equation}
\vec B_0=\gamma\Big(\vec B_{\perp} +{1 \over \gamma} \vec
B_{\parallel} + \sqrt{1-\gamma^{-2}} \Big[{\vec E_{\perp} \times
\vec n}\Big]\Big), \ \gamma = (1-\beta^2)^{-{1 \over 2}},
\end{equation}

\begin{equation}
\vec {M_0}=\vec {M}{_{0_{\parallel}}}+\vec {M_{0_{\perp}}},
\label{M_0}
\end{equation}
\begin{equation}
\begin{array}{c}
\displaystyle \vec {M}_{0_{\parallel}}=\gamma\vec\beta{n_{0} \over
\sqrt {1- v_{e}^{2}}}\left\{ \rho^{(1)}_{e}\left(1-{{\vec v}_e
\vec\beta \over {1- {\gamma^{-2}}}} \right)\right. \\-
\displaystyle\rho^{(2)}_{e}\left. \left(\vec\zeta_{e}\vec\beta
\sqrt{1-v^2_e}+ {(\vec \zeta_{e}{\vec v}_e)(\vec\beta{\vec v}_e)
\over 1+\sqrt{1-v^2_e} }\right){1 \over {1- {\gamma^{-2}}}}
\right\}, \label{M_0_parallel}
\end{array}
\end{equation}
\begin{equation}\label{M_0_perp}
\begin{array}{c}
\displaystyle \vec {M}_{0_{\perp}}=-\frac{n_{0}}{\sqrt {1-
v_{e}^{2}}}\Bigg\{ \vec{v}_{e_{\perp}}\Big(
\rho^{(1)}_{e}+\displaystyle\rho^{(2)}_{e}\frac
{(\vec{\zeta}_{e}{\vec{v}_e})} {1+\sqrt{1-v^2_e}}\Big)+
{\vec{\zeta}_{e_{\perp}}}\rho^{(2)}_{e}\sqrt{1-v^2_e}\Bigg\},
\end{array}
\end{equation}
where $t$ is time in the laboratory frame,  $\vec F_{\perp}$ and
$\vec F_{\parallel}$ ($\vec F= \vec B,\vec E$)  are transversal
and longitudinal (with respect to the direction $\vec n$ of
neutrino motion) electromagnetic field components in the
laboratory frame. For simplify  we neglect here the neutrino
electric dipole moment, $\epsilon=0$, and also consider the case
when matter is composed of only one type of fermions ( electrons).
The general case of $\epsilon\neq0$ and matter composed of
different types of leptons is discussed in our papers
\cite{GriLobStuPLB02, LobStuPLB03}. Here $n_0=n_{e}\sqrt
{1-v^{2}_{e}}$ is the invariant number density of matter given in
the reference frame for which the total speed of matter is zero.
The vectors $\vec v_e$, and $\vec \zeta_e \ (0\leqslant |\vec
\zeta_e |^2 \leqslant 1)$ denote, respectively, the speed of the
reference frame in which the mean momentum of matter (electrons)
is zero, and the mean value of the polarization vector of the
background electrons in the above mentioned reference frame. The
coefficients $\rho^{(1,2)}_e$ are calculated if the neutrino
Lagrangian is given, and  within the extended standard model
supplied with $SU(2)$-singlet right-handed neutrino $\nu_{R}$,
\begin{equation}\label{rho}
\rho^{(1)}_e={\tilde{G}_F \over {2\sqrt{2}\mu }}\,, \qquad
\rho^{(2)}_e =-{G_F \over {2\sqrt{2}\mu}}\,,
\end{equation}
where $\tilde{G}_{F}={G}_{F}(1+4\sin^2 \theta _W).$

\subsection{Neutrino spin oscillations in electromagnetic fields}

From the new generalized BMT equation \cite{EgoLobStuPLB00}  for
the neutrino spin evolution in an electromagnetic field and matter
(and also from the simplified version of the equation given by
(\ref{S})) the corresponding Hamiltonians describing neutrino spin
oscillations are just straightforward. Thus, {\it \bf the first
new effect} is the prediction of neutrino spin oscillations in
various electromagnetic field configurations. We have derived
\cite{EgoLobStuPLB00,DvoStuYF01_04, StuYF04} the new resonances in
neutrino oscillations for several electromagnetic fields such as
the field of  circular and linearly polarized electromagnetic
waves and superposition of an electromagnetic wave and constant
magnetic field. We have also studied (see the second paper of ref.
\cite{DvoStuYF01_04}) the possibility for parametric resonances in
neutrino oscillations in periodically varying electromagnetic
fields (an electromagnetic wave with varying amplitude and a
"castle wall" magnetic field of an undulator).

\subsection{Neutrino spin oscillations in matter and different
external fields}

We have predicted \cite{EgoLobStuPLB00,LobStuPLB01}, using the
generalized BMT equation, {\it \bf the second new effect}:
neutrino spin procession can be stimulated not only by presence of
electromagnetic fields (i.e., by electromagnetic interactions of
neutrino) but also by weak interactions of neutrino with
background matter. Thus, neutrino spin procession could appear
without any electromagnetic field in the presence of matter
through which neutrino propagates. This conclusion is just
straightforward also from the simplified version of the spin
evolution equation (\ref{S}).

Moreover, we have shown \cite{LobStuPLB03} that the neutrino spin
precession {\it always occurs} in presence of matter (even in the
case of non-moving and unpolarized matter) if the initial neutrino
state is not longitudinally polarized. Indeed, as it follows from
(\ref{S}), in the laboratory reference frame the corresponding
equation for the three-dimensional neutrino spin is
\begin{equation}\label{10}
{d{\vec S} \over dt} = 2\mu\rho^{(1)}_{e}n_{e}[\vec S \times \vec
\beta ].
\end{equation}
If neutrino is propagating along the OZ axis, $\vec \beta= (0,0,
\beta)$, then solutions of these equations for the neutrino spin
components are given by
\begin{equation}
S^{1}=S_{0}^{\perp} \cos \omega t,\, \,
 S^{2}=S_{0}^{\perp} \sin \omega t, \,\,
 S^{3}=S^{3}_{0}, S^{0}=S^{0}_{0},
 \end{equation}
where
\begin{equation}
\omega= 2\mu\rho^{(1)}_{e}n_{e}\beta,
\end{equation}
$S_{0}^{\perp}$ and $S^{3,0}_{0}$ are constants determined by the
initial conditions. Obviously, if $S_{0}^{\perp}\neq 0$ then the
solution of eq.(\ref{10}) is not trivial (non-zero) and there is a
procession of the neutrino spin vector $\vec S$ in the background
matter.

Within the developed Lorentz invariant approach it is also
possible \cite{DvoStuJHEP02} to find the solution for the neutrino
spin evolution problem in a more general case when  the neutrino
is subjected to general types of non-derivative interactions with
external fields that are given by the Lagrangian
\begin{equation}
-{\cal L}=g_{s}s(x){\bar \nu}\nu+ g_{p}{\pi}(x) {\bar
\nu}\gamma^{5}\nu+ g_{v}V^{\mu}(x){\bar \nu}\gamma_{\mu}\nu+
g_{a}A^{\mu}(x){\bar \nu}\gamma_{\mu}\gamma^{5}\nu+
{{g_{t}}\over{2}}T^{\mu\nu}{\bar \nu}\sigma_{\mu\nu}\nu+
{{g^{\prime}_{t}}\over{2}} \Pi^{\mu\nu}{\bar
\nu}\sigma_{\mu\nu}\gamma_{5}\nu,
 \end{equation} where $s, \pi,
V^{\mu}=(V^{0}, {\vec V}), A^{\mu}=(A^{0}, {\vec A}),
T_{\mu\nu}=({\vec a}, {\vec b}), \Pi_{\mu\nu}=({\vec c}, {\vec
d})$ are the scalar, pseudoscalar, vector, axial-vector, tensor,
pseudotensor fields, respectively.
For the neutrino spin evolution equation in this case we have
found
\begin{equation}\label{S_eq_gen}
\begin{array}{c}
\displaystyle {{d{\vec S} \over dt}}= 2g_{a}\left\{ A^{0}[{\vec
S}\times{\vec \beta}]- {{1}\over{1+{\gamma}^{-1}}} ({\vec A}{\vec
\beta})[{\vec S} \times{\vec \beta}]- {1 \over \gamma}[{\vec
S}\times{\vec A}] \right\}
\\ \displaystyle +2g_{t}\left\{ [{\vec S}\times{\vec b}]-
{{1}\over{1+{\gamma}^{-1}}} ({\vec \beta}{\vec b})[{\vec
S}\times{\vec \beta}]+
[{\vec S}\times[{\vec a}\times{\vec \beta}]] \right\} \\
\displaystyle + 2ig^{\prime}_{t}\left\{ [{\vec S}\times{\vec c}]-
{{1}\over{1+{\gamma}^{-1}}}({\vec \beta}{\vec c})[{\vec
S}\times{\vec \beta}]- [{\vec S}\times[{\vec d}\times{\vec
\beta}]] \right\}.
\end{array}
\end{equation}
It is worth to be noted that  (see also \cite{BerGroNarPRD99})
neither scalar nor pseudoscalar nor vector interaction contributes
to the neutrino spin evolution.

The neutrino spin evolution equation (\ref{S_eq_gen}) can be used
for any theoretical model in which neutrino has mentioned above
general interactions. For instance, within the standard model
 the weak interaction of a neutrino with the background matter has the axial vector term, whereas
the electromagnetic interaction of a neutrino is described by the
tensor field $T_{\mu\nu}=F_{\mu\nu}=({\vec E},{\vec B})$. As it
has been also recently shown \cite{DvoGriStu04}, the considered
general equation (\ref{S_eq_gen}) can be used for description of
the neutrino spin oscillations in a gravitational field of a
rotating object. In this case, the gravitational field in the
weak-field limit can be treated as an external axial vector field.

\subsection {Relativistic matter motion effects in neutrino
oscillations}

The derived new equation for the neutrino spin evolution enables
us to study spin oscillations in the case of moving with arbitrary
(also relativistic) speed and polarized matter. We have predicted
\cite{LobStuPLB01} {\it \bf the third effect}: the matter motion
can drastically change the neutrino oscillation pattern and, in
particular, can significantly shift the neutrino spin oscillation
resonance condition (see the third and fourth papers of ref.
\cite{SchVall81VolVysOku86LimMar88Akh88VidWud90Smi91AkhPetSmi93}),
if compared with the case of non-moving matter.

 We use again the simplified neutrino
spin evolution equation (\ref{S}). In the case of slowly moving
matter, $v_e\ll1$, from eqs. (\ref{M_0}), (\ref{M_0_parallel}),
and (\ref{M_0_perp}) we get
\begin{equation}
{\vec M}_0={n_e} \gamma{\vec\beta}\Big(
\rho^{(1)}_{e}-\rho^{(2)}_{e}\vec\zeta_{e}\vec\beta \Big),
\label{21}
\end{equation}
in agreement with results of \cite{Wol78,NunSemSmiValNPB97}. In
the opposite case of relativistic flux, $v_e\sim 1$, we find,
\begin{equation}
{\vec M}_0={{{n_0}} \over {\sqrt {1-v_{e}^{2}}}} \Big(
\rho^{(1)}_{e}+\rho^{(2)}_{e}\vec\zeta_{e}{\vec v}_e \Big) \Big(
1-\vec\beta{\vec v}_e \Big).
\end{equation}
One can easily see that the matter effect can be annihilated owing
to the relativistic motion of matter along the direction of
neutrino propagation, provided that $1-\vec \beta \vec
v_{e}\approx 0$. We also predict significant increase of matter
effect in neutrino spin oscillations for neutrino propagating
against the relativistic flux of matter.

To illustrate this phenomenon let us consider the case of neutrino
spin oscillations in the flux of electrons, that could move with
arbitrary (also relativistic) speed, under the influence of an
arbitrary constant magnetic field, $\vec B=\vec B_{\parallel}
+\vec B_{\perp}$. In the adiabatic approximation for the
particular case of electron neutrinos $\nu_e$ propagating in
matter composed of electrons, the probability of conversion $\nu_L
\rightarrow \nu_R$ can be written in the form \cite{LobStuPLB01},
\begin{equation}
P_{\nu_L \rightarrow \nu_R} (x)=\sin^{2} 2\theta_{eff}
\sin^{2}{\pi x \over L_{eff}},\end{equation}
\begin{equation}
sin^{2} 2\theta_{eff}={E^2_{eff} \over
{E^{2}_{eff}+\Delta^{2}_{eff}}}, L_{eff}={2\pi \over
\sqrt{E^{2}_{eff}+\Delta^{2}_{eff}}},
\end{equation}
where $E_{eff}=2\mu B_{\perp}$ (terms $\sim O(\gamma^{-1})$ are
omitted  here), and
\begin{equation}
\Delta_{eff}= V(1-\vec \beta \vec v_{e}) +{2\mu B_{\parallel}
\over \gamma}, \ \  V={G_F \over \sqrt{2}}{n_0 \over
\sqrt{1-v_{e}^{2}}}(1+4 \sin ^{2} \theta _{W}).
\end{equation}
As  it is mentioned above, the matter effect in $\Delta _{eff}$
can be "eaten" by the relativistic motion of matter if $(1-\vec
\beta \vec v_{e})\approx 0$. In the case of the neutrino and
matter relativistic motion ($\beta$ and $v_{e}\sim 1$) in opposite
directions ($1-\vec \beta \vec v_{e}\approx 2$), the matter term
contribution $V$ can be reasonably increased due to the presence
of a small term $\sqrt{1-v_{e}^{2}}$ in the dominator.

The analogous { \bf effect} also exist
\cite{LikStu95,GriLobStuPLB02}
 in neutrino flavour oscillations: the neutrino resonance
condition \cite{Wol78, MikSmi85} can be significantly modified if
matter is moving with relativistic speed. The probability of
neutrino conversion $\nu_e \rightarrow \nu_\mu$ in arbitrary
moving and polarized matter ( composed, for instance, only of
electrons) can be written in the form
\begin{equation}
P_{\nu_e \rightarrow \nu_\mu}(x)=\sin^{2} 2\theta_{eff} \sin^{2}
{\pi x \over L_{eff}}, \label{ver}
\end{equation}
where the effective mixing angle, $\theta_{eff}$, and the
effective oscillation length, $ L_{eff}$, are given by
\cite{GriLobStuPLB02}
\begin{equation}
\sin^{2} 2\theta_{eff}={\Delta^{2}\sin^{2} 2\theta \over
{\Big(\Delta \cos2\theta - A\Big)^2+ \Delta^{2}\sin^{2} 2\theta}},
L_{eff}= {2\pi \over {\sqrt {\Big(\Delta \cos2\theta - A\Big)^2+
\Delta^{2}\sin^{2} 2\theta}}}.
\end{equation}
Here $\Delta = {\delta m^{2}_\nu / {2 |\vec p\,|}}$ and $\vec p$
is the neutrino momentum, $\theta$ is the vacuum mixing angle and
\begin{equation}
\begin{array}{c}
A=\displaystyle\sqrt2 G_F{n_0 \over \sqrt{1-v^2_e}}\left\{(1-{\vec
\beta}{\vec v_e}) (1-\vec {\zeta}_e {\vec v_e})\right. +
\left.\displaystyle\sqrt{1-v^2_e} \left[ \vec {\zeta}_e {\vec
\beta} -{(\vec {\beta} {\vec v_e})(\vec {\zeta}_e {\vec v_e})
\over 1+\sqrt{1-v^2_e}} \right] \right\} . \label{A}
\end{array}
\end{equation}

One can see that the neutrino oscillation probability, $ P_{\nu_e
\rightarrow \nu_\mu}(x)$, the  mixing angle, $\theta_{eff}$, and
the oscillation length, $ L_{eff} $, exhibit dependence on the
total speed of electrons $\vec v_e$, correlation between $\vec
\beta$, $\vec v_e$ and polarization of matter $\vec \zeta_e$. The
resonance condition
\begin{equation}
 {\delta m^{2}_\nu \over {2 |\vec p\,|}}\cos 2\theta= A,
\label{res}
\end{equation}
at which the probability has unit amplitude, also depends on the
motion and polarization of matter and neutrino speed. It follows
that the relativistic motion of matter could provide appearance of
(destroy) the resonance in the neutrino oscillations in certain
cases when for the given neutrino characteristics, $\delta
m^{2}_\nu$, $|\vec p\,|$ and $\theta$, and the invariant matter
density at rest, $n_0$, the resonance is impossible (exists).
A detailed analysis of the neutrino effective potential in moving
and polarized matter for different particular cases ( for
different speeds and polarizations of matter) can be found in
\cite {GriLobStuPLB02}.

\subsection{Spin light of neutrino in background environments}

The {\it \bf  fourth new effect} in neutrino oscillations is the
prediction \cite{LobStuPLB03,DvoGriStu04} for the new mechanisms
of electromagnetic radiation by neutrino moving in background
matter and/or electromagnetic and gravitational fields. The new
mechanism of electromagnetic radiation ( we have named
\cite{LobStuPLB03} this radiation \emph{"spin light of neutrino"}
($SL\nu$) ) originates from the neutrino spin precession that can
be produced whether by weak interactions with matter or by
interactions with external electromagnetic fields
\cite{LobStuPLB03}, or by interactions with gravitational fields
\cite{DvoGriStu04}. This radiation in the case when the neutrino
spin precession is induced by a constant magnetic field was also
considered before in \cite{BorZhukTernSPJ88}.

The total power of the $SL\nu$ does not washed out even when the
emitted photon refractive index in the background matter is equal
to unit. That is why the $SL\nu$ can not be considered as the
neutrino Cerenkov radiation (see, for example, \cite{IoaRaf97} and
references therein).

If we assume, for definiteness, that the spin light radiation is
produced by the electron neutrino $\nu_{e}$ moving in unpolarized
($\zeta_e=0$) matter composed of only electrons ( the more general
cases are also considered in \cite{LobStuPLB03}) and constant
magnetic field $\vec B=\vec B_{\perp} +\vec B_{\parallel}$ then
for the total spin light radiation power we get
\begin{equation}\label{BG}
I_{SL\nu}={64 \over 3} \mu^{6} \gamma^{4}\Big[ \big(
n_{e}\rho^{(1)}_{e} \vec \beta (1- \vec \beta \vec v_e) - {1 \over
\gamma} n_{e} \rho^{(1)}_{e} \vec v_{e_\perp}\big)^{2}+\vec B_{
\perp} + {1 \over \gamma} \vec B_{
\parallel}  \Big]^{2},
\end{equation}
where the terms proportional to ${\gamma^{-2}}$ in the brackets
are neglected. The spin light emission rate in non-moving matter
and in the absence of the magnetic field is
\begin{equation}\label{W}
\Gamma_{SL\nu}={\sqrt{2} \over
3}\gamma^{2}\mu^{2}G_{F}^{3}n^{3}_{e}.
\end{equation}
The $SL\nu$ in matter must be important for environments with high
effective densities, $n$, because the total radiation power is
proportional to $n^{4}$. The total power of the $SL\nu$ is
increasing with the neutrino energy increase and is proportional
to the fourth power of the neutrino Lorentz factor, $I_{SL\nu}
\sim {\gamma }^{4}$. It has been also shown \cite{LobStuPLB03}
that the $SL\nu$ is strongly beamed in the direction of neutrino
propagation and is confined within a small cone given by
$\delta\theta\sim\gamma^{-1}$.

The average energy of photons of the spin light in matter is
\begin{equation}
\omega_{SL\nu} = \sqrt {2} G_{F} n_{e} \gamma^{2}.
\end{equation}
For the density $n_{e}\geq 10^{30} cm^{-3}$ and neutrino with mass
$m_{\nu}=1 \ eV$ and energy $p_{0}=10 \ MeV$ the energy range of
emitted photons could span up to gamma-rays. These properties of
the $SL\nu$ in matter enables us to predict that this radiation
should be important in different astrophysical environments
(quasars, gamma-ray bursts etc) and in dense plasma of the early
Universe.

Let us briefly discuss the properties the $SL\nu$ produced by the
neutrino spin evolution in gravitational fields like that of
rotating neutron stars and black holes. If we consider
\cite{DvoGriStu04} the neutrino propagating along the rotation
axis of a massive object then the photon energy (in the laboratory
frame) of the $SL\nu$  in the gravitational field of an object can
be estimated as
\begin{equation}\label{omega}
  \omega_{SL\nu} \sim\omega_0 \gamma \sim
  \frac{G_{N}L}{r^3}\gamma^2,
\end{equation}
where $L$ is the  angular momentum of a rotating object, and $G_N$
is the Newton's constant. If the angular momentum is chosen to be
equal to the maximal allowed value $L=r_0^2/(4G_N)$ ( see, for
instance, \cite{Lan_Lif_Nauka88}),  where $r_0$ is the
Schwarzschild radius, then for the photon energy we have
\begin{equation}\label{omegavalue}
  \omega_{SL\nu}\sim 10^{-11}\times \gamma^2
  \left(
  \frac{r_0}{r}
  \right)^3
  \thinspace\text{eV}.
\end{equation}
It follows that $\omega_{SL\nu}\sim 10\thinspace\text{GeV}$ for
$\gamma \sim 10^{12}$ and $r \sim 10 \ r_0$. Note that for the
mass of neutrino $m_\nu\sim 1\thinspace\text{eV}$ the condition
$(\omega/p_0)\sim 10^{-2}\ll 1$ is still valid, i.e. the
quasiclassical approach to the neutrino spin evolution can be
applied in this case.

It should be noted that the whole developed approach to the
neutrino spin evolution and oscillations in different background
environments ( including the predictions for the corresponding new
effects) can be generalized, with a minor modification, for
neutrinos with the flavour changing magnetic moments. These cases
correspond to models of the Dirac neutrinos with non-diagonal
magnetic moments or the Majorana neutrinos which could also have
transitional (magnetic) moments.

\section{Conclusion}
We have developed the Lorentz invariant approach to the neutrino
flavour and spin oscillations and on this basis studied {\it \bf
the four new effects}: i) the neutrino spin oscillations in
various electromagnetic field configurations, ii) the possibility
for neutrino spin oscillations to be produced by weak interactions
with matter and by gravitational fields of rotating massive
objects, iii) the significant change in the neutrino (spin and
flavour) oscillations pattern due to the relativistic motion of
matter, and iv) the spin light radiation by a neutrino moving in
matter and/or electromagnetic and gravitational fields (the {\it
spin light of neutrino}, $SL\nu$, in background environments). We
believe that {\it \bf the four new effects} could have important
consequences in different astrophysical and cosmological
environments.

I should like to thank Francois Vannucci and Daniel Vignaud for
the invitation to participate in the XXI International Conference
on Neutrino Physics and Astrophysics and for support of my stay in
Paris.

\end{document}